\documentclass[a4paper,reqno,12pt,draft]{article}
\usepackage{amssymb,euscript,bbold}

\newtheorem{thm}{Theorem}[section]
\newtheorem{conj}{Conjecture}
\newtheorem{hyp}{Hypothesis}
\newcommand{\be}{\begin{equation}}
\newcommand{\ee}{\end{equation}}
\newcommand{\eq}[1]{Eq.~(\ref{eq:#1})}
\newcommand{\ba}{\begin{eqnarray}}
\newcommand{\ea}{\end{eqnarray}}

\newcommand{\ft}{\footnote}

\newcommand{\TeV}{{\rm TeV}}
\newcommand{\vol}{{\rm vol~}}

\newcommand{\p}{\partial}

\newcommand{\BZ}{\mathbb{Z}}

\begin{document}
\input{epsf}

\begin{flushright}
IC/2006/042
\end{flushright}
\begin{flushright}
\end{flushright}
\begin{center}
\Large{\sc A Finite Landscape?}\\
\bigskip
\bigskip
\large{\sc B.S. Acharya}
{\renewcommand{\thefootnote}{}
\footnotetext{bacharya at ictp.it, mrd at physics.rutgers.edu}}\\
\bigskip\large
{\sf Abdus Salam\\ International Centre for Theoretical
Physics,\\
Strada Costiera 11\\ 34014 Trieste. Italy}\\
\bigskip\large{\sc M.R. Douglas}\\
\bigskip\large
{\sf New High Energy Theory Center,\\
Rutgers University,\\
Piscataway, New Jersey}\\
and\\
{\sf I.H.E.S., Bures-sur-Yvette, France}
\end{center}
\bigskip
\begin{center}
{\bf {\sc Abstract}}
We present evidence that the number of string/$M$ theory vacua
consistent with experiments is finite. We do this both
by explicit analysis of infinite sequences of vacua and by
applying various mathematical finiteness theorems.

\end{center}


\newpage

\section{Introduction}
\normalsize
\bigskip

How many string vacua are there which agree with experimental
observations?  While obviously a difficult question, perhaps using
statistical arguments which take advantage of the large number of
vacua, we can estimate this number.  If it is small, this will
demonstrate that string theory is falsifiable in principle, and help us
decide what type of predictions the theory can make.

A first step in this direction would be to establish whether or not
this number is finite.  One simplification of this is that we might
only need a subset of the properties of the Standard Model to
establish finiteness.  The basic properties we assume are four large
dimensions and no massless scalar fields.

In thinking about this question and attempting to
construct infinite sequences of vacua we were led to the following
conjecture (slightly refined from that made in \cite{stat}):

\begin{conj}\label{con}

The number of 4d vacua with an upper bound on the vacuum energy,
an upper bound on the compactification volume,
and a lower bound on the mass of the lightest Kaluza-Klein tower, is finite.

\end{conj}

We will discuss the precise definition of the quantities which enter
this conjecture below.  Their simplest motivation is phenomenological
and they can all be related to observables in four dimensions.

The only compactifications we really understand well enough at present
to use in studying this conjecture are those which are based on large
volume limits; {\it i.e.} begin with compactification of the $d=11$
and $d=10$ supergravity theories, and then add various stringy or $M$
theoretic effects to stabilize the moduli.  
On general grounds infinite sequences of such vacua might emerge in
three different ways:

\begin{description}
\item{a) {\it topology}}: there could exist infinitely many topologies
for the extra dimensions, or for the vector bundles used in defining
gauge fields in the extra dimensions.

\item{b) {\it fields}}: there could exist infinitely many distinct solutions
of the equations of motion, including Einstein metrics,
brane configurations or background gauge fields.  
A basic example here is to add an arbitrary number $N$ of space-filling
brane-antibrane pairs.  We suspect that such configurations are always
unstable for sufficiently large $N$, and discuss this in section 
\ref{sec:energy}.  But we haven't proven it, so our conjecture
explicitly postulates an upper bound on the vacuum energy.  Such a
bound is easy to motivate phenomenologically, as the actual vacuum
energy is very small.

Among more subtle results for this question,
in \cite{Douglas:2004yv} it was pointed out that theorems in algebraic
geometry bound the allowed values of generation number and the number
of branches of bundle moduli space in heterotic string
compactification, while in \cite{Douglas:2006xy} it was recently shown
that intersecting brane models on the $T^6/\BZ_2\times\BZ_2$
orientifold are finite in number.

\item{c) {\it fluxes}}: an infinite number of fluxes.  This is
conceptually not so different from (b), but we separate it out as it
can be studied more systematically.  By now there are a fair number of
results showing that flux vacua are finite in number; let us mention
results in \cite{Eguchi} and a theorem proven in
\cite{DougLu}
according to which the integrals of the densities on
Calabi-Yau moduli space, which give the asymptotic number of flux
vacua in IIb and other compactifications, are finite. 

\end{description}

In the next section we will describe sequences of vacua with
infinitely many topologies and explain why they do not violate our
finiteness conjecture.  We go on to describe Cheeger's finiteness
theorem in Riemannian geometry \cite{cheeger}, and how it can be
applied to demonstrate that certain classes of string vacua (eg
Calabi-Yau and $G_2$ vacua) contain only finitely many topological
choices for the extra dimensions in the supergravity approximation.
This establishes that, even if there were infinitely many
topologically distinct Calabi-Yau or $G_2$ manifolds, only finitely
many of them can be consistent with the Standard Model.  In the last
section of the paper we address the question of infinitely many field
configurations, giving an example of an infinite sequence which does
not contradict \ref{con}; we finish by relating the finiteness of the
number of vacua to Gromov's compactness result \cite{gromov} for the
space of Riemannian metrics..

The fact that experimental observations put a bound on the allowed
topologies of the extra dimensions is, whilst striking from a physical
point of view, a consequence of the interplay between `dynamics' and
topology in the context of Riemannian geometry.

\section{Infinite topologies?}

Are there infinitely many topologically distinct choices for the extra
dimensions in string/$M$ theory? The answer is yes, as we will
illustrate by example shortly.  However, we will also demonstrate that
in many classes of vacua in the supergravity approximation, {\it only
finitely many topological choices can be consistent with experimental
observations}.  In particular, it is not known whether or not there
are finitely or infinitely many diffeomorphism types of compact
Calabi-Yau threefolds or $G_2$-holonomy manifolds. Our results will
show that {\it it does not matter} that the number could be
infinite, by proving there can be at most finitely many vacua which
are consistent with the conjecture.

Before describing this result, we will begin by discussing 
explicit examples of infinite sequences of topologies for the
extra dimensions. These examples originate in the fact that there
exist infinite sequences of topologically distinct Einstein
manifolds with positive cosmological constant.  For example, consider
quotients
of round spheres by cyclic groups of {\it arbitrary} order. Since in
dimension seven such Einstein manifolds form part of the data for
Freund-Rubin vacua \cite{fr} (see \cite{duff} for a review), we have
infinite sequences of topologies for the extra dimensions in such
vacua. In this section we will explicitly analyse these and more
non-trivial infinite sequences of Einstein manifolds and show how they
are consistent with the conjecture.

\subsection{Review of Freund-Rubin vacua}

These are near horizon geometries of
brane metrics in which there is an $AdS$ factor. We
will focus on the $M$2-brane case for which the
Freund-Rubin metric takes the form
\be\label{eq:frmetric}
g_{10+1} = g(AdS4_a) + a^2 g^{(0)}(X)
\ee
Here $AdS4_a$ is anti de Sitter space of radius $a$ and
$X$ is a compact 7-manifold.  We take non-zero electric four-form flux
$F\propto\epsilon_{0123}$; then Einstein's equation 
\be\label{eq:einstein}
R_{ij} = {1 \over 3} F_{ipqr} F_j^{pqr} - {g_{ij} \over 36}(|F|^2 - m^2)
\ee
forces  $g_0(X)$ to be an Einstein metric.  
We define this to have fixed positive scalar curvature, satisfying
\be\label{eq:fixcurve}
R_{ij}(g^{(0)} ) = 6 g^{(0)}_{ij}.
\ee
Note that this condition determines the overall 
normalization of $g^{(0)}$, and thus we call it a ``normalized metric.''

Since the actual metric
on the seven extra dimensions is rescaled by $a^2$, the
scalar curvature of $X$ is of order $a^{-2}$, as is that
of the four dimensional AdS universe.
Because of this, if the volume of $X$ as measured by
$g_0 (X)$ is order one,
then the masses of Kaluza-Klein modes are of order the
gravitational mass in AdS. In this case, there would be no
meaningful four dimensional limit in which one can ignore the
dynamics of the Kaluza-Klein particles.
While this is true for $S^7$, if we could find an example in which
the radii of $X$ in the $g^{(0)}$ metric were much smaller than one, there
would be a mass gap between the gravitational fluctuations
and Kaluza-Klein modes, and there would be a
four dimensional limit \cite{frrevisited}.

\subsubsection{Quantization of $a$}

If we think of the metric as the near horizon geometry of a brane
metric, then the parameter
$a$ is related to the integer number $N$ of branes, as
\be\label{eq:aquant}
a^6 = {N L_p^6\over V_0(X)},
\ee
where $L_p=1/M_p$ is the 
eleven dimensional Planck length and $V_0$ is the volume of
$X$ as measured by $g_0(X)$.

To show that \eq{aquant} is correct, we need the formulae for $g(AdS4)$
and the $G$-flux in the vacuum. These are respectively
\be
g(AdS4_a) = {a^2 \over r^2}dr^2 + {r^4 \over a^4}g_{2+1}
\ee
where $g_{2+1}$ is the $2+1$-dimensional Minkowski metric, and
\be
G \sim a~ dVol(AdS4_a) = {r^5 \over a^4} dV_{2+1}\wedge dr
\ee
where $dV_{2+1}$ is the volume form on the Minkowski space.

Because this is the background solution for $N$ $M$2-branes,
the $M$2-brane charge, which is the integral of $*G$ over $X$ 
in the metric \eq{frmetric}, must be $N$.
Hence
\be\label{eq:compN}
N = \int_X *G = a^6 V_0
\ee
We prefer to write physical quantities in terms of $N$ rather than
$a$.

The four dimensional Planck scale $M_{pl4}$ is related to the 
volume of $X$ in 11d Planck units as
\be \label{eq:fourMpl}
{M_{pl4}^2 \over M_p^2} = V(X) \sim \frac{N^{7 \over 6}}{V_0^{1 \over 6}} 
\ee
Notice that as $V_0$
{\it decreases}, $V$ actually {\it increases}. This may at first seem
counterintuitive, but actually it is a simple
consequence of the quantization of flux.
In order to maintain the integrality of \eq{compN}, a decrease in
$V_0$ must be accompanied by an increase in $a$ and thus $V$. 
This fact will play an important role in what follows.

The (negative) cosmological constant is of order
\be
|\Lambda| \sim M_{pl4}^4 {V_0^{\frac{1}{2}} \over N^{\frac{3}{2}}}
\ee

The above two formulae can also be rewritten as: 
\be V^3 |\Lambda|
= N^2 \ee and \be V^9 |\Lambda|^7 = V_0^2 \ee

\subsection{Finite number of solutions}

Let us begin by considering $X=S^7$.  From the relations above, we see
that as $N$ tends to infinity, the physical volume goes to
infinity. Thus, Freund-Rubin vacua are simple examples of infinite
sequences of vacua which do not violate the finiteness conjecture;
rather they violate one of its assumptions.

What if we consider more general compactification manifolds?  Since
the relation \eq{fourMpl} between $V_0$ and the four dimensional
Planck scale followed directly from \eq{compN}, which is true
regardless of the topology of the extra dimensions, we can make the
same argument for any infinite sequence of vacua in which the
normalized volumes approach zero, to show that the actual volume of
the extra dimensions will again go to infinity.

A simple example is the infinite sequence of Einstein 7-manifolds
constructed by taking a ${\bf Z_k}$ $\subset U(1)$ quotient of an
Einstein manifold with $U(1)$-symmetry.   Here $V_0 = (V_0(k=1))/k$ by
symmetry.

What about the opposite possibility, a sequence $X_i$ in which
the normalized volumes go to infinity?  This is not possible, because of 
\begin{thm} (Bishop, 1963 \cite{Berger})
A $d$-dimensional manifold $X$ with a lower bound on the Ricci tensor,
\be\label{eq:Riccibound}
R_{ij} \ge (d-1)k g_{ij}
\ee
with $k>0$, has volume less than or equal to that of the
round ($SO(d+1)$-symmetric) sphere of curvature $k$, 
$$
\vol M \le \vol S^d(k),
$$
with equality only for $X\cong S^d$ (Cheng 1975).
\end{thm}
Our normalized metrics satisfy \eq{Riccibound} with $k=1$, so
at fixed flux $N$, all of the others lead to larger physical volumes
$V$ than the round sphere.

Thus, to invalidate the conjecture with generalized Freund-Rubin
vacua, we must find an infinite sequence of Einstein metrics with
finite normalized volume.  Now there do exist much more non-trivial
infinite sequences of distinct Einstein metrics. For instance,
\cite{boyer} construct several infinite sequences of Einstein
7-manifolds by constructing sequences of conical eight dimensional
hyperkahler quotients. The bases of these cones are Einstein
7-manifolds. Unfortunately, to our knowledge the volumes have not been
computed in these examples.  We will argue later that they cannot
provide counterexamples, but this will follow from more abstract
considerations; here let us turn to a more explicit case.

Recently in the context of supergravity solutions and the AdS/CFT
correspondence, Gauntlett {\it et al} \cite{gauntlett} have constructed
infinite sequences of Einstein 7-metrics, again with positive scalar
curvature. The volumes of these examples have been computed in
unpublished work of Martelli and Sparks \cite{martellisparks}.  These
Einstein 7-manifolds $Y^{p,k}(M)$ are labelled by two integers $(p,k)$
and are $S^3/{\bf Z_p}$ bundles over Kahler Einstein 4-manifolds $M$.
The two classes of explicit examples are when $M$ is either ${\bf
CP^2}$ or ${\bf CP^1}{\times}{\bf CP^1}$.  More 
recently, \cite{cvetic, martelli} have shown that these $Y^{p.k}$
metrics are special cases of more general families of Einstein
metrics which depend upon three integers.

Let us discuss the manifolds $Y^{p,k}({\bf CP^2})$;
the others are very similar.
The integer $k$ is bounded by the values of $p$ as \be 3p/2 \leq k
\leq 3p. \ee
Thus, to obtain an infinite sequence we must take $p$
to infinity. The most important result of \cite{gauntlett} for our
purposes is the nature of the manifold for the `boundary values'
of $k$. One can show that \be Y^{p,3p}({\bf CP^2}) = {S^7}/{\bf
Z_{3p}} \ee and that \be Y^{p,3p/2}({\bf CP^2}) = {M^{3,2}}/{\bf
Z_{p/2}} \ee where $M^{3,2}$ is the homogeneous Stiefel
manifold.

Thus the two limiting values of $k$ are cyclic quotients of
homogeneous Einstein manifolds. To proceed further we need some
information about the volumes of $Y^{p,k}({\bf CP^2})$. One can
show (cf \cite{martellisparks}) that 
\be\label{eq:ineqvol} 
V_0({M^{3,2}}/{\bf Z_{p/2}}) >
 V_0 (Y^{p,k}({\bf CP^2})) > V_0 ({S^7}/{\bf Z_{3p}}) \ee

{}From these bounds it follows that as $p$ tends to infinity $V_0
(Y^{p,k}({\bf CP^2}))$ goes to zero. Hence by our previous
discussion the actual volume of the extra dimensions diverges
owing to the flux quantization condition.

To summarise: we have shown in several explicit infinite sequences
of four dimensional Freund-Rubin vacua that only finitely many
can satisfy an upper bound for the volume of the extra dimensions.

\section{Cheeger's Finiteness Theorem}

A remarkable theorem of Cheeger can help explain such results.

\begin{thm} (Cheeger, 1970  \cite{cheeger}):\\
\smallskip
In a (potentially infinite) sequence of Riemannian manifolds $M_i$
with metrics
such that \\1. the sectional curvatures $K$ are all bounded, say $|K|
\leq 1$\\
2. the
volumes are bounded below; $V_i \geq V_{min}$\\
3. the diameters are bounded above, $D_i \leq D_{max}$ ; \\
\smallskip
there can only be a finite number of diffeomorphism types.
\end{thm}

In less formal language, this theorem states that given that all
components of the Riemannian curvature tensors (in an orthonormal
basis) of all the manifolds in the sequence remain bounded, and that
the manifolds do not become too elongated in some direction, and that
their volumes do not go to zero, then there can be only finitely many
topological types in the sequence.

Let us first explain how we can use this theorem to establish our
conjecture, in the supergravity approximation, in a wide variety of
cases.  We will then explain the mathematics behind it in an intuitive
way.  Note that the conditions of the theorem involve dimensionful
quantities.  We will take these to be measured in units of the 
largest fundamental length scale (the string scale or the 10 or
11-dimensional Planck scale).

We first want to
argue that the hypotheses of Conjecture \ref{con} imply the
hypotheses of Cheeger's theorem.  First, conditions 1 and 2 are
trivial in the following sense.  In general, string and $M$ theory physics
can be described using supergravity and Riemannian
geometry in the ``supergravity regime,'' in which curvatures and
other field strengths are small
compared to the string or higher dimensional Planck scales, 
all closed geodesics are longer than the string scale,
volumes of minimal cycles are larger than these scales, and so on.
Thus, we must impose conditions 1 and 2 with $V_{min}=1$,
just to remain in this regime.

Of course, one knows that string/$M$ theory is sensible beyond this regime,
and thus our finiteness arguments cannot cover all possible vacua.
The point however is that we want to exclude violations of our
conjecture in this regime, a problem which can be addressed using
present-day mathematics.  If it holds there, we can go on to think
about other, more stringy regimes later.

This leaves condition 3.  This condition bounds the diameter $D$,
defined as the supremum of the distance between any pair of points.
Intuitively, one can argue that as the diameter becomes large, the
manifold elongates and the smallest eigenvalues of the Laplacian will
go to zero, so that a Kaluza-Klein mode becomes light.  Thus, a lower
bound on the KK scale, should imply an upper bound on the diameter.
Phenomenological bounds on the KK scale will be around $1 \TeV$ (for
particle with Standard Model quantum numbers), or $(10 \mu m)^{-1}\sim
10^{-3} {\rm eV}$ for graviton KK modes, which would lead to
corrections to the inverse square law for gravity.

To make this precise, we begin by considering the smallest non-zero eigenvalue
$\lambda_1$ of the scalar Laplacian on the extra dimensions, 
\be \Delta\phi = \lambda_1 \phi .
\ee 
Writing 
\be
g_{\mu\nu}^{(d+4)} = g_{\mu\nu}^{(4)} \cdot \phi + \ldots,
\ee
we see that this will be the mass squared of a graviton KK mode in four
dimensions.  

Next, by taking powers $\phi^n$ of the wave function, we might expect
to get approximate eigenfunctions with eigenvalues $n\lambda_1$, and
an entire Kaluza-Klein tower.  However this is not obvious; a single
field might be becoming light, which could happen in various ways
unrelated to the diameter condition.  Nevertheless, let us
provisionally take $\lambda_1$ as the definition of the Kaluza-Klein
mass squared, and return to this point below.

We want to argue that, if we put a phenomenological lower bound on $\lambda_1$,
\be\label{eq:MKKbound}
M_{KK}^2 \le \lambda_1,
\ee
this will enforce an upper bound on the diameter, justifying condition 3.
This will be true if we can find a mathematical
upper bound on $\lambda_1$ in terms of the diameter, say of the general form 
\be\label{eq:Lambound}
\lambda_1 \le \frac{C}{D^2}
\ee
suggested by dimensional analysis,
where $C$ might depend on other data in some bounded way.  Combining
this with the phenomenological bound \eq{MKKbound}, we would have
the desired upper bound,
$$
D^2 \le \frac{C}{M_{KK}^2} .
$$

In fact a bound \eq{Lambound} can be proven by variational arguments
\cite{Chavel}.  We will explain the details of this shortly, but first
let us discuss what this will imply.  We have now argued that, if we
restrict attention to compactifications in the supergravity regime, a
sequence of vacua satisfying the hypotheses of conjecture \ref{con}
will come from a sequence of compactification manifolds satisfying the
hypotheses of Cheeger's theorem.  Therefore, the sequence can only
contain a finite number of distinct topologies.  In particular the
sequence could be ``all'' vacua satisfying our conditions, so to get
all vacua we would only need to consider a finite list of possible
topologies, even if there turn out to be infinitely many topologies of
(say) Calabi-Yau manifolds.  All but a finite number of these would 
lead to vacua with unobserved light KK modes.

Why should Cheeger's theorem be true?  The basic intuition can perhaps
be seen by thinking of a Riemann surface of constant negative
curvature.  To increase the genus, one must attach handles; this
always increases the volume (due to the simplicity of two dimensions)
but does not ``obviously'' increase the diameter.  However, if one
thinks of the Riemann surface as some piece of the Poincar\'e disk
with identifications along its boundaries, it is clear that increasing
the volume must increase the diameter, albeit at a very slow
(logarithmic) rate.

The actual proof is rather intricate but breaks down into several
components.  The basic idea is to use the hypotheses to show that any
manifold $M_i$ in the sequence can be covered by a definite finite
number $N$ of convex balls of a fixed radius $r$, and use this to
reduce the problem to combinatorics.  To show this, one combines
various theorems in ``comparison geometry,'' according to which the
local geometry of any manifold satisfying curvature bounds has to be
``similar'' (in some precise sense) to a constant curvature space.
This allows bounding the minimal volume of a ball, which is clearly
necessary.  Subtle arguments are needed to show that one can actually
use balls of a definite radius; in particular one might worry that the
manifolds might contain very short noncontractable loops, which would
be a problem.  A lower bound on the length of such loops is the {\it
injectivity radius} $r_i$, defined as the minimum over all points $p$
of the radius at which Riemann normal coordinates around $p$ break
down (are no longer one-to-one).  This will be the shorter of half the
length of the shortest periodic geodesic, or else the shortest
distance between conjugate points.  It turns out that the injectivity
radius can be bounded below, and that doing this requires imposing an upper
bound on the diameter, rather than the volume or something else.

Once we have shown that $M_i$ can be covered by a finite number of
balls, we can imagine describing the topology of $M_i$ as a simplicial
complex, in which balls become vertices, a pairwise overlap becomes a
link, a triple overlap becomes a face, and so on.  Because the balls
are convex, the overlap regions are all contractible, so this complex
will have the same homotopy type as the original manifold.  In
addition, the curvature conditions can be used to show that the
transition functions between the balls are diffeomorphisms, so the
complex actually determines the diffeomorphism type.  Then, the
evident fact that given a finite number of vertices, one can make only
a finite (though very large) number of different simplicial complexes,
implies the theorem.

Now let us come back to \eq{Lambound}.  The basic intuition here is
simple; one chooses two points $p$ and $q$ whose separation 
$d(p,q)$ is the diameter $D$, and considers a wavefunction
with derivatives of order $\pi/D$,
$$
\phi(x) = \sin \frac{\pi d(p,x)}{D} - {\rm const}.
$$
where the constant is chosen to make it orthogonal to the ground state,
$0=\int \phi$.
A variational argument should then tell us that
\be
\lambda_1 \le \frac{\int \phi\Delta\phi}{\int \phi^2} \sim \pi^2/D^2.
\ee
This is basically right but there is an important caveat which emerges
from making a proof along these lines.  The standard proof
\cite{Chavel} is made
by dividing the manifold into two pieces with boundary, 
$M_p$ within distance $D/2$
of $p$ and $M_q$ within distance $D/2$ of $q$.  
Assume we have a lower bound on the Ricci curvature:
$$
R_{ij} \ge (d-1)k g_{ij}
$$
for some real constant $k$.  One can then show 
using a variational argument that the first nontrivial eigenvalues
on each of the two pieces with Dirichlet boundary conditions are 
bounded above,
\be\label{eq:lambdaonebound}
\lambda_1 \le \lambda_1|_{M_p}, \lambda_1|_{M_q} \le \lambda_1(k,D/2) ,
\ee
by the Dirichlet eigenvalue on a ball of radius $D/2$
and constant curvature $(d-1)k$.
The smaller of these is then an upper bound for $\lambda_1$ on $M$.

A similar argument can be used to bound the $m$'th eigenvalue, by
dividing $M$ into $m+1$ regions.  One finds
$$
\lambda_m \le \lambda_1(k,\frac{D}{2m})
$$
and thus $\lambda_m \sim m^2/D^2$ as is appropriate for a KK tower.
This justifies the definition we made, in the sense that while the true
$\lambda_1$ might not be the first mode of a KK tower, the $\lambda_1$
which is bounded by \eq{lambdaonebound} will be the first of a tower.

The eigenvalue on the ball $\lambda_1(k,D/2)$ can be written
explicitly in terms of Bessel functions.  For non-negative Ricci
curvature, it indeed falls off as $1/D^2$.  On the other hand,
large curvature can also make the lowest non-zero eigenvalue large,
as one can see from the ball; for negative $k<<-1/D^2$ we have
$\lambda(k,D/2)\sim (d-1)^2|k|/4$.

This caveat is not important for the Freund-Rubin case
(actually Bishop's theorem forces $k\sim 1/D^2$ for positive
curvature anyways), so we have now justified the claims made in 
section 2, and can assert that the number of topologies which can
lead to quasi-realistic vacua is finite.  This is not the end of the
story as we can imagine infinitely many solutions on a single
topology, but we discuss this in section 4.

The caveat is also unimportant for Ricci flat compactification
manifolds such as Calabi-Yau and $G_2$ holonomy, so we can make the
same statement there:
even if there were an infinite number of Calabi-Yaus or
$G_2$-holonomy manifolds, only finitely many topologies can be
used as models for the extra dimensions in the supergravity
approximation.

What about the case of negative Ricci curvature?  One might at first
think that this would be ruled out by some sort of positive energy
theorem, but this is not so obvious; for example the trace term in
Einstein's equations \eq{einstein} can be negative in the presence of
magnetic flux.

We should distinguish two cases.  On the one hand, more realistic
solutions will have branes, flux, quantum effects and so on,
potentially leading to corrections to the metric and small amounts of
negative Ricci curvature.  This will not significantly affect the
spectrum of the Laplacian and the bound \eq{lambdaonebound}, so is not
a problem for the argument.

On the other hand, if we could find supergravity solutions with
compactification manifolds with large negative Ricci curvature
compared to $1/D^2$, we might well find that the Laplacian has a gap
independent of the diameter.  While Cheeger's theorem would still tell
us that any infinite series of compactifications must run off to
infinite diameter, if the volume and the KK scale stay bounded, there
might be no sign of this from four dimensional physics.

This seems on the face of it  unlikely, and indeed the flux
contributions to the stress-energy will fall off with compactification
volume.  Actually, we do not know of static supergravity solutions
with negative curvature, so perhaps there is a general argument 
against this possibility.

\subsection{Singularities}

Cheeger's theorem applies to smooth manifolds.  In string theory
however, special kinds of singularities are physically acceptable. For
example orbifold singularities and conical singularities in space are
often physically sensible.  Is there a natural extension of Cheeger's
theorem which includes singularities of certain kinds?

In fact for 4-dimensional spaces there is such a generalisation, due
to Anderson \cite{anderson}.  Anderson proves that the set of compact
four dimensional orbifolds in which the magnitude of the Ricci tensor
is bounded, the volume is bounded below and the diameter above
contains finitely many topological types. In other words, if one
replaces condition 1 of Cheeger's theorem with a bound on the Ricci
tensor and allows orbifold singularities then there are again finitely
many topological types. From a physical point of view the bound on the
Ricci tensor is much more natural since this is equivalent to a bound
on the energy-momentum tensor which is completely reasonable.
Unfortunately, extensions of Anderson's result to higher dimensions
have only been obtained with an additional bound on the $L^{n/2}$ norm
of the Riemann tensor, which does not appear to have a clear
physical interpretation. This latter condition implies that the
curvature of the extra dimensions cannot diverge too quickly at the
singularities.  It would be interesting to see, if by including more
general conical singularities, a more ``physically'' reasonable
version of Anderson's theorem applies.

\subsection{Infinite sequences of non-supersymmetric vacua}
\label{sec:energy}

One can also motivate other ways to set the bounds in the theorem.

For non-supersymmetric vacua, we could try to argue for
condition 1, the curvature bound, on grounds
of stability. Let us consider the fluctuations of the
metric $g_{ij}(X)$ in the extra dimensions. The modes $\delta g_{ij}$
corresponding to scalar fields in four dimensions have a contribution
to their mass-squared given by eigenvalues of the Lichnerowicz
operator on $X$,
\be\label{eq:Lich}
 -\nabla_i^2 \delta g_{ij} +
2R_i^k \delta g_{jk} -2 R_{i\;\;\;j}^{\;\;m\;\;n} \delta g_{mn} \equiv
\Delta_L \delta g_{ij} \propto m^2 \delta g_{ij} 
\ee
Unlike the Laplacian, $\Delta_L$ can have negative eigenvalues
on a compact manifold, leading to possible tachyons and instability.
This is of course classical and
small tachyonic masses might be compensated for by quantum
corrections, but large Riemann curvatures will lead to large tree level
masses which cannot be compensated.
Thus, stability suggests the imposition of the first condition.

How generic a problem is this?  For supersymmetric compactifications,
a scalar mass squared will be related to a fermion mass $m_F$ as 
\be
m^2 =  m_F(m_F-\sqrt{3|\Lambda|}/M_{pl4}) .  
\ee 
For Minkowski compactifications, this is manifestly non-negative,
while more generally minimizing with respect to $m_F$ leads to the
Breitenlohner-Freedman bound.  For phenomenology, we are most interested
in either the physical case $\Lambda\sim 0$, or the case of 
$\Lambda=-3|W|^2 << M_{pl4}^4$ so that adding small supersymmetry
breaking effects will bring $\Lambda$ to zero.  Either way, there is
a small window $0<m_F<\sqrt{3|\Lambda|}/M_{pl4}$ for instability, while large
variations of the Riemann tensor in \eq{Lich} would presumably lead
to large variations of $m_F$ and push $m^2$ positive again.  Thus
this general type of instability due to KK modes would be controlled,
as long as the supersymmetry breaking scale is below the Planck scale.

More generally, this type of consideration motivates placing an upper
bound on the supersymmetry breaking scale.  By the standard supergravity
formula
\be
V = |F|^2 + |D|^2 - 3|W|^2 ,
\ee
this will imply an upper bound on the vacuum energy.

\section{Infinite classes of solutions}

One must next ask whether there can exist infinitely many distinct
vacuum metrics for a fixed topology.\ft{We particularly thank Is
Singer for emphasizing this issue.}  For instance, infinitely many
Einstein metrics on a fixed topology.  In the Calabi-Yau and
$G_2$-holonomy cases the answer is of course yes, there is a
continuous moduli space of metrics. In these cases however, this
degeneracy is removed by fluxes and/or other contributions to the
moduli potential, and finiteness in the physically relevant case of
vacua without massless fields is addressed by studying flux vacua.

In the Freund-Rubin case, for instance, the situation is
different. Einstein metrics of positive scalar curvature tend to be
rigid. However, for a fixed topology there can exist infinitely many
distinct Einstein metrics!  To study this concretely, we need an
explicit infinite sequence for which the volumes are known. In
dimension 7 we do not know of such examples, however there does
exist an infinite series of Einstein metrics on $S^2 \times S^3$,
which were also discovered in \cite{gauntlett} and depend upon two
integers $(p,q)$.  Analogously to equation \eq{ineqvol}, one can show
that the volumes of these manifolds are bounded above and below by the
volumes of ${\bf Z_p}$ quotients of homogeneous spaces
\cite{mart}. This shows that only finitely many of these manifolds
have a finite volume. It would be interesting to find out whether this
type of result generalizes to any infinite sequence of Einstein metrics with
positive cosmological constant on a space with fixed topology.

\subsection{Finiteness, barriers and distances between vacua}

The examples of continuous Calabi-Yau and $G_2$ moduli spaces show
that we will in any case need to bring in more than Riemannian
geometry and Einstein's equations to proceed.  Since the full problem
of moduli stabilization is very complicated, we again ask what
simplifications could be made which would still suffice to get a
convincing argument for finiteness, and an estimate of the number of
solutions.

One observation of this type \cite{stat} is that, if we know in some
example that a potential is generated (say by nonperturbative
effects), we can estimate the number of vacua by counting the number
of minima of a ``generic'' potential.  For the example of
supergravity, this suggests that the number of vacua is the integral
of a natural characteristic class for the bundle in which the 
covariant gradient of the superpotential $D_iW$ takes values, and
this argument leads to the Ashok-Douglas density \cite{AD}.

Let us make a simpler observation of this type.  Suppose we can
formulate a problem of finding stabilized vacua in terms of a
potential $V(\phi)$ which is a function of some scalar fields
$\phi^a$.  These fields might have been moduli in a related theory
with more supersymmetry, or not; the main point however is that
a four-dimensional effective action containing these fields,
$$
\int d^4x G_{ab}(\phi) \partial \phi^a\partial\phi^b - V(\phi) +
\ldots
$$
will define a metric $G_{ab}$ on the configuration space parameterized
by $\phi$.  Now computing this metric exactly is probably even more
difficult than computing the potential.  But qualitative properties of
the metric which imply finiteness might not be so hard to get.

A reasonable conjecture, made in various talks of the second author
around 2004--2005, is that the finite number of minima of $V(\phi)$ in
some region of configuration space is tied to the finiteness of the
volume of that region, in the volume form $\sqrt{\det G}$.  This
suggests the general conjecture 
\cite{HorneMoore,Douglas:2005hq,Vafa:2005ui}
that the volumes of all of the moduli spaces which arise naturally
in string compactification (Calabi-Yau moduli space, moduli spaces of
conformal field theories, and so on) are finite.  Of course just saying
this is not in itself a physical argument for why such
volumes should be finite; other arguments which have been proposed
have been based on the renormalization group in CFT 
\cite{Douglas:2005hq}, or consistency after incorporating gravity
\cite{Vafa:2005ui}.  

However, one can easily imagine loopholes to a connection between
numbers of vacua, and the volume of configuration space.
Mathematically, the basic example is the function
$$
V(\phi) = \sin \frac{1}{\phi}
$$
which has an infinite sequence of minima accumulating at zero.
If we take the canonical kinetic term for $\phi$, there are infinitely
many minima in a region of finite volume.

Of course this is just postulating a function and there is no reason
to think that such a potential can arise physically.  On the other
hand we would like some criterion which distinguishes it from the
potentials we think can arise physically, to see why we should not
worry about this possibility.\footnote{
A somewhat more physical model of this type was proposed in
\cite{Dvali:2003br}, but so far as we know it does not come out of
string theory compactification.}

One such is to ask that the distance between vacua in field space, be 
comparable to the energy scale $(\Delta V)^{1/4}$ set by the height of
the potential barrier between the vacua.  This is certainly a natural
condition if the physics does not depend on additional scales; whether 
or not it is universally true, we do not know.  In any case
there are toy potentials satisfying this condition 
with an infinite number of vacua, for example
\be\label{eq:toyV}
V(\phi) = \phi^8 \sin \frac{1}{\phi} .
\ee
However such potentials will (by definition) lead to sequences of
vacua for which the successive barrier heights go to zero.  Thus, if
we were working at any fixed energy scale $E$, as we approached
$\phi\rightarrow 0$ we would eventually decide that the potential was
becoming unimportant.  This suggests a definition in which,
if two vacua are only separated
by a potential barrier less than some prespecified minimal height 
$\epsilon$, we should consider them as the same physical vacuum.

A related but slightly different criterion would be to insist that
physically distinct vacua have some minimal separation in
configuration space.  One reason to prefer this definition is that
vacua with different vacuum expectation values for fields will in
general make different predictions even if not separated by a
potential barrier, as is familiar for compactifications with extended
supersymmetry (consider masses of BPS states).
Conversely, if two minima of $V(\phi)$ are separated
by a distance less than some prescribed minimal separation $\epsilon$ in
configuration space, we might consider them as the same physical vacuum.
One might argue that in this case quantum fluctuations will produce
transitions between the different vacua.  

Another argument for this definition would be to think about the 
physical observables as functions of the field values in configuration
space: say masses, couplings, etc. are all written as
$$
m(\phi), g(\phi), \ldots
$$ 
To the extent that these are continuous functions, there will be
some $\Delta\phi$ below which we cannot distinguish the physical
predictions.  This argument also suggests its own limitations -- if
there are phase transitions, i.e. other parameters in the
configuration which we neglected, which after a small shift in $\phi$
can vary by large amounts, this need not be true.  So we get an idea of a
``sufficient'' specification of the configuration, i.e. one in which
we included all the fields which could vary significantly and which
affect the observables.

The upshot of this rather general discussion is the following
\begin{hyp}
There exists a minimal distance $\epsilon$ in configuration space
between physically distinct vacua.
\end{hyp}
Its justification comes from two alternatives.  The simpler is that
this is true of the actual potentials which arise from string
theory.  But, although we do not know of examples, one can imagine 
potentials such as \eq{toyV} which describe
infinite series of vacua with accumulation points.  If these lead to
distinct observable physics, then Conjecture \ref{con} will be wrong.
On the other hand, if the physical observables also converge in the
limit, we should count vacua which are separated by distance
less than $\epsilon$ as physically ``the same,'' also justifying the
hypothesis.

\subsection{Convergence and precompactness}

One reason we bring this topic up is that in the mathematics
literature we have been citing, a rather central idea is that of a
topology on the space of Riemannian manifolds.  In other words, one
wants to be able to say when a sequence of manifolds $M_i$ converges,
and if so, to what.

This can be related to the idea of distance on configuration space
we were just discussing.  Suppose we have a family of metrics for
the compactification manifold $X$ with parameters $\phi^i$; then
the parameters can be interpreted as fields in four dimensions,
the Einstein (or supergravity) action specialized to the family can
be interpreted as a potential 
$$
V(\phi) = \int_X \sqrt{g}R[g(\phi)] ,
$$
and the metric on the space of metrics will give the kinetic term,
$$
G_{ab}(\phi) =
 \int_X \sqrt{g} g^{ij}g^{jl}
 \frac{\p g_{ij}}{\p\phi^a}
 \frac{\p g_{kl}}{\p\phi^b} .
$$
In particular, the metric on configuration space will imply a
topology on the configuration space.

Of course topology is a weaker concept than metric; if we modify
our definition of distance in any continuous way, we will get the
same topology.  In fact it is really too weak to make any statement
analogous to ``finiteness of volume.''  For example, the interval
$\phi\in(0,1)$ which contained an infinite number of vacua for 
the potential \eq{toyV}, is topologically equivalent, say under the
map $\phi\rightarrow \phi/(1-\phi)$, to the entire
positive real axis $\phi\in(0,\infty)$.  We would not be surprised to
learn that the latter
configuration space contained an infinite number of vacua.

However the mathematicians work with something in between topology and 
metric, which we now explain.  Suppose we have a distance function
$d(x,y)$ 
between pairs of points (say points in moduli space).
Such a function (satisfying simple axioms such as positivity and
the triangle inequality) is the general notion of metric, as opposed to
a Riemannian metric.
We then define convergence in terms of this as
$$
\lim_{i\rightarrow\infty} x_i= x\ {\rm iff} \lim d(x_i,x)=0.
$$
Now suppose we had another metric $d'(x,y)$, with
$$
d'(x,y) = \lambda(x,y) d(x,y) \qquad \forall x,y.
$$
If the $\lambda$'s are bounded away from zero and infinity,
we will get the same idea of convergence.  Thus such an equivalence class
of distance functions incorporates the topology of the underlying
space.  And it carries more information, for example $d(x,y)=|x-y|$ 
on the interval $(0,1)$ is
not equivalent in this sense to $d(x,y)=|1/x-1/y|$ on $(0,1)$.

Now, given this structure, we can talk about a set $X$ having
``finite diameter'' or ``infinite diameter,'' according to whether the set
of pairwise distances 
$$
\{d(x,y): x,y\in X\}
$$
is bounded or not.  Rather than finite diameter, one usually speaks of
$X$ being {\it totally bounded}.  This is true if $X$ can be covered
by a finite number of sets, each of finite diameter.  This includes 
the case in which $X$ has more than one connected component.

One can prove that this is equivalent to the condition
that $X$ is {\it precompact}.  
As the name suggests, this means that $X$ can be Cauchy completed
(as in Cauchy's construction of the real numbers) and the resulting
space is compact (every sequence contains a convergent subsequence).

It is tempting to add to the earlier conjecture, that a finite volume
configuration space can only contain a finite number of vacua, 
the conjecture that a precompact configuration space can only contain
a finite number of vacua.  Now if we accept the hypothesis of the last
subsection, this is not a conjecture but in fact follows from the
definitions, since regions of configuration space of diameter less than
$\epsilon$ can contain at most one vacuum.

We now quote
\begin{thm} (Gromov 1981 \cite{gromov})\\

Let ${\cal M}_{d,D,k}$ be the space of $d$-dimensional
Riemannian manifolds with diameter not larger than $D$,
and ${\rm Ricci} \ge (d-1)k$.

This space is precompact in  the Gromov-Hausdorff metric. 
\end{thm}

This is a configuration space which describes all possible metrics
(satisfying the conditions), including the Einstein metrics and all
other metrics.  If we believe the notion of distance given by the
Gromov-Hausdorff metric has any relation to physical distance, and
accept the arguments of the previous subsection, this shows that there
can only be a finite number of vacua including not just all
topologies, but all solutions to the Einstein equations as well.

The Gromov-Hausdorff metric is not very physical.  A better
model for the dependence of physical observables on the underlying 
manifold is the spectral geometry approach of B\'erard, Besson and
Gallot \cite{BBG}.  As briefly summarized in \cite{Solvay},
BBG define a metric between manifolds in terms of the
distances between the entire set of eigenfunctions of the Laplacian.
So, convergence in this metric corresponds much more directly to
convergence of observables, the spectrum and wave function
overlaps.  One could also use this to show that this form of convergence
agrees with the one defined using distances in the moduli space metric when
that makes sense (this one is more general).

BBG then show that ${\cal M}_{d,D,k}$ is precompact using this notion
of distance.  So, there are at most a finite number of
compactifications in this sense.  Admittedly, it is hard to imagine
that such a general result (which does not even impose the equations
of motion) will lead to a useful bound on the number, but this
illustrates the idea.

Kontsevich and Soibelman 
have suggested that this type of definition can be made
for conformal field theories as well, and have conjectured
\begin{conj} (Kontsevich and Soibelman 2000 \cite{KontSoi})\\

Let ${\cal M}_{c,\Delta}$ be the space of two-dimensional conformal
field theories with central charge $c$, and with no operators of 
dimension less than $\Delta$ (except the identity).
This space is precompact in a ``natural'' metric.
\end{conj}
The diameter condition is replaced by a lower bound on operator
dimensions using the sort of relation we gave in section 3; Kontsevich
has argued that CFT's always satisfy a version of ``${\rm Ricci}\ge
0$'' avoiding the subtlety discussed there.  Note also that there is
no volume condition; T-duality suggests that there is no limit in
which the volume can go to zero.  Clearly it would be very interesting
to make a definition of the space of CFT's and develop this; see
\cite{Roggenkamp:2003qp} for work in this direction.

Another corollary of these theorems is that any infinite distance
limit in a moduli space of metrics is associated to a violation of the
hypotheses; restricting attention to the supergravity regime, it must
be associated to the diameter of the manifold going to infinity.  By
our discussion in section 3, this implies that a tower of KK modes is
becoming light.  This is ``Conjecture 2'' in the recent
\cite{OoguriVafa}; we see that in the case of sigma models in the
large volume limit it follows from known facts in geometry, while for CFT's
it would follow from Kontsevich and Soibelman's conjecture.

{\sf Acknowledgements.}
An early version of this work was presented by the second author
at Strings 2005 in Toronto, and both authors thank the organizers
of that meeting for hospitality there.  MRD particularly thanks
Maxim Kontsevich and Alex Nabutovsky for inspiring discussions; 
BSA is especially grateful to Dario Martelli and James Sparks for 
explaining \cite{martellisparks}.
We also thank Frederik Denef, Greg Moore, Sameer Murthy, Eva Silverstein, Is Singer
and Cumrun Vafa for discussions.
The work of MRD is supported in part by DOE grant DE-FG02-96ER40959.


\begin{thebibliography}{10}

\bibitem{stat}
  M.~R.~Douglas,
  JHEP {\bf 0305} (2003) 046
  [arXiv:hep-th/0303194].

\bibitem{Douglas:2004yv}
  M.~R.~Douglas and C.~G.~Zhou,
  JHEP {\bf 0406}, 014 (2004)
  [arXiv:hep-th/0403018].

\bibitem{Douglas:2006xy}
  M.~R.~Douglas and W.~Taylor,
  arXiv:hep-th/0606109.

\bibitem{Eguchi}
  T.~Eguchi and Y.~Tachikawa,
  JHEP {\bf 0601}, 100 (2006)
  [arXiv:hep-th/0510061].

\bibitem{DougLu}
  M.~R.~Douglas and Z.~Lu,
  ``On the Geometry of Moduli Space of Polarized Calabi-Yau
  manifolds,''
  math.DG/0603414.

\bibitem{cheeger} J. Cheeger, 
``Finiteness theorems for Riemannian Manifolds''
Amer. J. Math. {\bf 92} (1970) 61.

\bibitem{gromov} M. Gromov, ``Structures metriques pour les varietes Riemanniennes,''
CEDIC, Paris 1981, Eds. J. Lafontaine and P. Pansu.

\bibitem{fr}
  P.~G.~O.~Freund and M.~A.~Rubin,
  Phys.\ Lett.\ B {\bf 97} (1980) 233.

\bibitem{duff}
  M.~J.~Duff, B.~E.~W.~Nilsson and C.~N.~Pope,
  Phys.\ Rept.\  {\bf 130}, 1 (1986).

\bibitem{frrevisited}
  B.~S.~Acharya, F.~Denef, C.~Hofman and N.~Lambert,
  arXiv:hep-th/0308046.

\bibitem{Berger} M. Berger, ``A Panoramic View of Riemannian Geometry,''
Springer 2003.

\bibitem{boyer} C. Boyer, K. Galicki, B. Mann, E. Rees, 
``Compact $3$-Sasakian $7$-manifolds with arbitrary second Betti number.''
Invent. Math. 131 (1998), no. 2, 321--344.

\bibitem{gauntlett}
  J.~P.~Gauntlett, D.~Martelli, J.~F.~Sparks and D.~Waldram,
  Adv.\ Theor.\ Math.\ Phys.\  {\bf 8}, 987 (2006)
  [arXiv:hep-th/0403038].

\bibitem{cvetic}
  M.~Cvetic, H.~Lu, D.~N.~Page and C.~N.~Pope,
  Phys.\ Rev.\ Lett.\  {\bf 95}, 071101 (2005)
  [arXiv:hep-th/0504225].

\bibitem{martelli}
  D.~Martelli and J.~Sparks,
  Phys.\ Lett.\ B {\bf 621}, 208 (2005)
  [arXiv:hep-th/0505027].

\bibitem{martellisparks} D. Martelli and J. Sparks, unpublished notes.

\bibitem{mart}
  D.~Martelli and J.~Sparks,
  Commun.\ Math.\ Phys.\  {\bf 262}, 51 (2006)
  [arXiv:hep-th/0411238].

\bibitem{Chavel} I. Chavel, ``Eigenvalues in Riemannian Geometry,''
Academic Press 1984.

\bibitem{anderson} M. Anderson, 
``Convergence and Rigidity Under Ricci Curvature Bounds,''
Invet. Math. 102 (1990) 429.

\bibitem{g2mod}
  B.~S.~Acharya,
  arXiv:hep-th/0212294.

\bibitem{ADV}
  B.~S.~Acharya, F.~Denef and R.~Valandro,
  JHEP {\bf 0506}, 056 (2005)
  [arXiv:hep-th/0502060].



\bibitem{AD}
  S.~Ashok and M.~R.~Douglas,
  JHEP {\bf 0401} (2004) 060
  [arXiv:hep-th/0307049].

\bibitem{deWolfe}
  O.~DeWolfe, A.~Giryavets, S.~Kachru and W.~Taylor,
  JHEP {\bf 0507}, 066 (2005)
  [arXiv:hep-th/0505160].



\bibitem{HorneMoore}
 J.~H.~Horne and G.~W.~Moore,
  Nucl.\ Phys.\ B {\bf 432}, 109 (1994)
  [arXiv:hep-th/9403058].



\bibitem{Solvay}
  M.~R.~Douglas,
  arXiv:hep-th/0602266.

\bibitem{Douglas:2005hq}
  M.~R.~Douglas and Z.~Lu,
  arXiv:hep-th/0509224.


\bibitem{Vafa:2005ui}
  C.~Vafa,
  arXiv:hep-th/0509212.

\bibitem{Dvali:2003br}
  G.~Dvali and A.~Vilenkin,
  Phys.\ Rev.\ D {\bf 70}, 063501 (2004)
  [arXiv:hep-th/0304043].

\bibitem{BBG} P. Berard, G. Besson, S. Gallot, ``Embedding Manifolds by Their Heat Kernel,''
Geom. Funct. Anal. {\bf 4} (1994) 373.

\bibitem{KontSoi}
M. Kontsevich and Y.~ Soibelman,
 [arXiv:math.SG/0011041]

\bibitem{Roggenkamp:2003qp}
  D.~Roggenkamp and K.~Wendland,
  Commun.\ Math.\ Phys.\  {\bf 251}, 589 (2004)
  [arXiv:hep-th/0308143].

\bibitem{OoguriVafa}
 H.~Ooguri and C.~Vafa, [arXiv:hep-th/0605264].

\end{thebibliography}
\end{document}